\RequirePackage{fix-cm}
\documentclass[twocolumn]{svjour3}          
\smartqed  
\usepackage{amssymb}

\usepackage{graphicx}
\usepackage{hyperref}
\usepackage{amsmath}
\usepackage{cite}
\usepackage{color}
\usepackage{appendix} 
\usepackage{flushend}
\usepackage{ulem}
\def\Tau{{\mathcal T}}
\def\makeheadbox{\relax}

\begin{document}

\title{Performance of a temporally multiplexed single-photon source \\with imperfect devices}

\author{Agustina G. Magnoni \and Ignacio H. L\'opez~Grande \and Laura T. Knoll \and M.A. Larotonda}

\institute{Agustina G. Magnoni \and Ignacio H. L\'opez~Grande \and Laura T. Knoll \and M.A. Larotonda
                \at DEILAP-UNIDEF, CITEDEF-CONICET, J.B. de La Salle 4397, 1603 Villa Martelli,
                      Buenos Aires, Argentina             
             \and 
            Agustina G. Magnoni \and Ignacio H. L\'opez~Grande \and Laura T. Knoll \and M.A. Larotonda
                \at Departamento de F\'{\i}sica, FCEyN, UBA. Ciudad Universitaria, 1428 Buenos Aires, Argentina, \\
            \email{mlarotonda@citedef.gob.ar}
}

\date{}

\maketitle

\begin{abstract}
Scalable photonic quantum technologies require highly efficient sources of single photons on demand. Although much progress has been done in the field within the last decade, the requirements impose stringent conditions on the efficiency of such devices. One of the most promising approaches is to multiplex a single or several heralded photon sources into temporal modes. In this work we analyze a specific proposal to synchronize photons from a continuous source with an external reference clock using imperfect optical switches, which necessarily degrade the ideal behavior of the devised arrangement. The performance of the source as a sub-poissonian light emitter is studied taking into account losses in the multiplexing arrangement, detector efficiency and dark counts. We estimate a fivefold increase in the single photon probability achieved for 0.5 dB loss switches. 
\end{abstract}

\section{Introduction}

One of the key requirements for building a quantum computing device, a quantum processor, or even to physically realize a quantum communication protocol, is a system that can encode quantum information. The optical photon is an attractive physical system for encoding a quantum bit, essentially due to its ability to be guided long distances with low losses, and to be controlled by (optical) linear operations. Whether the quantum processing is performed with photons or with another technology, the transmission of quantum information between different devices must be carried out with photonic quantum bits. 

In order to take full benefit of photons as quantum information carriers, any implementation of a quantum processing task requires a single photon source, linear and nonlinear operation-performing components and photon counting devices. In recent years, significant progress has been achieved in the development of fast and efficient single photon detection \cite{natarajan2012superconducting,degen2017quantum} and quantum photonic gates \cite{carolan2015universal,ribeiro2016demonstration,volz2014nonlinear,costanzo2017measurement}. 
Regarding the single photon source, the usual approach of using attenuated light pulses has several drawbacks that limit the performance of the implemented quantum algorithms. An efficient and scalable single photon source suitable for quantum information processing is yet to be engineered. Thus, a great deal of research is currently in progress to obtain devices that deliver light pulses carrying a single photon in well-defined spatio-temporal and polarization modes. Different approaches include the use of single-emitter sources and Spontaneous Parametric Down Conversion (SPDC).

Single emitter sources include many devices based on fluorescence from atoms or ions \cite{kimble1977photon,mandel1979sub,diedrich1987nonclassical}, molecules \cite{basche1992photon} and quantum dot-like structures in solid state environments.
Early results have been obtained with self-assembled III-V and II-VI semiconductors\cite{holmes2014room,sebald2002single,couteau2004correlated}, and with epitaxially growth InGaAs and GaAs quantum dots \cite{michler2000quantum,gammon1996homogeneous}. Recently, an enhancement on brightness and purity has been obtained by coupling the quantum dots with electrically controlled cavities, in deterministically
fabricated devices \cite{somaschi2016near}.  Single photon emission from Nitrogen-vacancy centers in a diamond lattice and other solid-state single photon emitters have also been demonstrated and extensively studied, and reviewed in \cite{aharonovich2016solid}.

On the other hand, the most efficient and widespread type of single photon sources are the ones based on nonlinear frequency processes such as SPDC, where a pair of photons is produced in a material with a non-zero second order susceptibility $\chi^{(2)}$. One of the photons is sent to a detector, which heralds with high probability the presence of the other photon, provided the two downconverted modes are non degenerated in some degree of freedom. State-of-the-art implementations involve integrated optics devices \cite{bock2016highly} and heralding efficiencies exceeding 50\% \cite{montaut2017high}.

A critical issue in down-conversion is the fact that, even though the heralding process does indeed remove the zero photon component of the heralded field ($P_0$), there is still a non-negligible probability of generating more than one pair. This probability, due to the thermal nature of the generated field, scales with pump intensity. In order to keep the multiphoton emission low, parametric sources become inherently inefficient, particularly in applications that require many single, simultaneous photons. A solution to these scaling problems are the multiplexed single photon sources. They include either an array of heralded sources and/or an arrangement of active delay lines and a switching network to build an approximation to a true ``photon gun'', or  on-demand single photon source. These sources rely on the combination of  spatial \cite{collins2013integrated,mazzarella2013asymmetric,christ2012limits,ma2011experimental,jennewein2011single,shapiro2007demand,migdall2002tailoring,adam2014optimization} or temporal \cite{adam2014optimization,mendoza2016active,schmiegelow2014multiplexing,glebov2013deterministic,mower2011efficient,jeffrey2004towards,zhang2017indistinguishable,rohde2015multiplexed,Kwiat_old,kwiat_new} multiplexing of one or several SPDC sources and active switching to tailor the photon statistics of the output pulses, in order to increase the single photon output probability ($P_1$) without increasing the  multiple photon pulse emission. 
Spatial multiplexing schemes rapidly become unpractical, since the number of photon sources and heralding detectors scales with the amount of multiplexed modes. Instead, temporal multiplexing uses a single detector and photon source, thus significantly increasing the efficiency and scalability. Schemes presented in \cite{mower2011efficient,schmiegelow2014multiplexing} require an electronic timing circuit to reference detection times from the heralding photons to an external clock. This temporal reference is then used to control a switching network that actively modifies the delay imposed to the heralded photon, coupling it into a temporal mode defined by the clock.

Real-world implementations, however, impose limitations on the efficiency and, most important, on the fidelity of the quantum output state of these engineered sources. Best-in-class $2\times2$ fast, piezoceramic electro-optical switches impose an insertion loss of the order of 0.9 dB \cite{xiong2016active}, preventing arbitrarily large variable delay networks. Also, the heralding detector non-unity efficiency and spurious counts add uncertainty to the heralding process. A correct assessment on the capacity of these actively controlled multiplexing sources to produce high quality single photon pulses (or at least to deliver pulses with a sub-poissonian photon statistics) is of utmost importance for any potential application. These include mature technologies such as Quantum Key Distribution \cite{heindel2012quantum,schiavon2016heralded} and mid and long term developments in large scale optical quantum computers and quantum simulators \cite{hayes2010fault,stace2011loss,gimeno2015three,harris2018linear}.

In this work, we study in detail the expected photon statistics produced by a temporally Multiplexed Single Photon Source (MSPS) that was originally introduced in \cite{schmiegelow2014multiplexing} based on bulk components consisting of a periodically poled nonlinear crystal as a single heralded photon source and a network of fiber optic components. We obtain  a simplified expression for the ideal photon output distribution,  and we include the effect of non-ideal components (such as component losses, detector dark counts, detector efficiency). The effect of lossy components in alternative multiplexed-source architectures has been studied in \cite{bonneau2015effect} and \cite{adam2014optimization}. The first one is based on arrays of heralded single photon sources (spatial multiplexing) and the last one presents an analysis of both spatial and time multiplexing. 

A theoretical model for the photon distribution is presented, together with a computational simulation of the complete experiment. We first analyze the behaviour of the source under ideal conditions and finally we include real life implementation losses. It is worth to note that the source generates time-synchronized photons (i.e. pulsed output) using a CW pump.

\section{Multiplexed single photon source}
\label{sec:multiplex}

The specific MSPS addressed here relies on a scheme that produces clock synchronized single photons from one SPDC source, using one of the two generated photons as a herald. Initially, the time difference between a clock and the detections of the heralding photons is measured. This determines the amount of delay that must be imposed to the heralded photons in order that they exit the source with their wave-packet temporally synchronized with the next clock tick. An equivalent  arrangement was experimentally demonstrated in 2016 \cite{xiong2016active}, and constitutes a promising result for scalable multiplexing of non-deterministic photon sources towards a single near-deterministic source.

\begin{figure}[h]
  \includegraphics[width=0.48\textwidth]{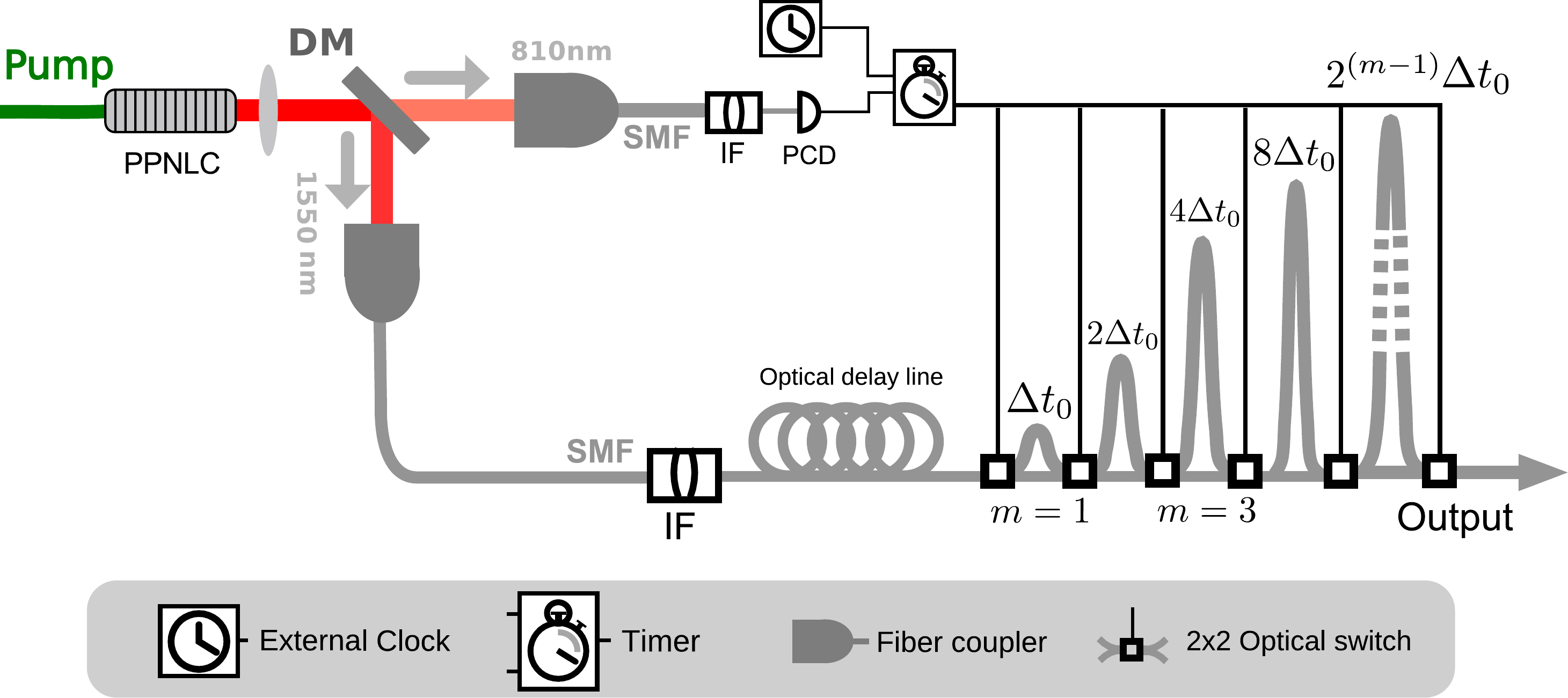}
  \caption{Schematic view of the time-multiplexed single photon source considered in this work. A pair of photons is generated at a random time by spontaneous parametric down-conversion in a periodically poled nonlinear crystal (PPNLC) and separated using a dichroic mirror (DM). Both paths are coupled to single mode fibers (SMF). The idler photon is detected with a photon counting device (PCD) and used as a herald. The time difference between this detection and a tick of an external clock is used to alter the path of the signal photon so that it arrives at the output synchronized with the clock tick. Path modification is done by a series of fast fiber switches connected to a binary-division fiber network. Also, interferometric Fabry-Perot filters  (IF) limit the downconverted photons bandwidth, ensuring coherence times long enough to match the precision of the time measurement and correction.}
  \label{fig:sketch}
  \end{figure}

After a detection of the herald photon within a temporal window, defined as a binary fraction of the clock period, the heralded photon is re-routed to the adequate delay line (by actively switching different temporal segments or \textit{correction stages}), and always leaves the system synchronized with the clock tick. The maximum temporal offset that can be corrected and the precision of this synchronization scheme depends on the amount of correcting stages and the size of the shortest delay, respectively. 
As a consequence, the source has two main components: the correlated photon pair generation, where the heralding (\emph{idler}) and the heralded (\emph{signal}) photons are created; and the synchronization of the signal photon with the external clock tick. This last operation is performed using a binary division arrangement, which minimizes the passages through switches. A sketch of the source is depicted in figure \ref{fig:sketch}.

The multiplexing system is based on a chain of optical delays, designed following a binary-division strategy. The $m$ stages in the chain are connected via fast optical switches, and each stage may or may not add a selected delay. The delay of the first stage, $\Delta t_0$, is matched to the coherence length of the downconverted photons.
A convenient setting for this temporal window is $\Delta t_0=2$~ns, which is a tradeoff between the intensity loss due to spectral narrowing (this condition implies a spectral narrowing of the idler photon beam down to a bandwidth of $\Delta \lambda \approx 4$~pm), and the temporal precision imposed by the detector and electronic jitter \cite{schmiegelow2014multiplexing}.

Each of the subsequent correcting stages include a delay line that doubles the length of the preceding one. The maximum delay that can be implemented with $m$ correction stages is $(2^m -1) \Delta t_0$. Taking into account the option of bypassing all the delays, the period of the synchronizing clock is $\Tau = 2^m \Delta t_0$.

The efficacy of the setup as a single photon source relies on the fact that there is a high probability of detecting a photon during the whole temporal window or total synchronization interval $\Tau$, while there is a low probability of occurrence of multiphoton events within a single temporal window $\Delta t_0$. These probabilities depend on the input mean photon number $\mu$, of the SPDC source. 

The photon distribution at the exit of the MSPS ($\mathcal{P}_n$), in an ideal situation, can be expressed as follows:

\begin{equation}
\label{ec:p1_ideal}
\mathcal{P}_n(\mu ) = \left\{
  \begin{array}{lr}
    P_0(\mu _T) & : n = 0\\
    (1- P_{0}(\mu _T))\frac{P_n (\mu)}{(1-P_0(\mu ))}  & : n \neq 0
  \end{array}
\right.
,
\end{equation}
where $\mu$ is the input mean photon number per detection window $\Delta t_0$, for a heralding photon rate $r$: $\mu = \Delta t_0\times r$. Additionally, we define $\mu _T$ as the mean photon number for the whole temporal window ($\mu _T = 2^m \mu $). $P_n$ is the Poisson probability of obtaining $n$ photons ($e^{-\mu}\frac{\mu^n}{n!}$), for the respective mean photon number $\mu$. 
When at least one pair of photons is emitted by the source within the whole temporal window (the probability of this is given by $1-P_{0}(\mu _T)$), the heralding and re-routing process is on. The probability of having a $n$-photons state output synchronized with the clock tick is the Poisson coefficient for a single temporal window, normalized by a factor that takes into account that there cannot be zero photons within the window.

In order to increase the probability of single photon emission of the device, the input mean photon rate $\mu$ can be tuned by changing the pump power to maximize the single photon output probability, $\mathcal{P}_n(\mu )$. 
We can define an ``optimum input photon rate'' ($\mu_{opt}$), which maximizes the single photon output probability:
\begin{equation}
\label{ec:mu_opt}
\mu_{opt}=\arg\!\max_{\mu}\mathcal{P}_1(\mu).
\end{equation}

As the amount of correction stages increases, the maximum probability of single photon output is obtained with smaller values of $\mu $. For $\mu < \mu _{opt}$ and $\mu > \mu _{opt}$, the dominant condition is zero photon output and multiple photon output, respectively \cite{schmiegelow2014multiplexing}.

\begin{figure}[h!]
\centering
\includegraphics[width=0.5\textwidth]{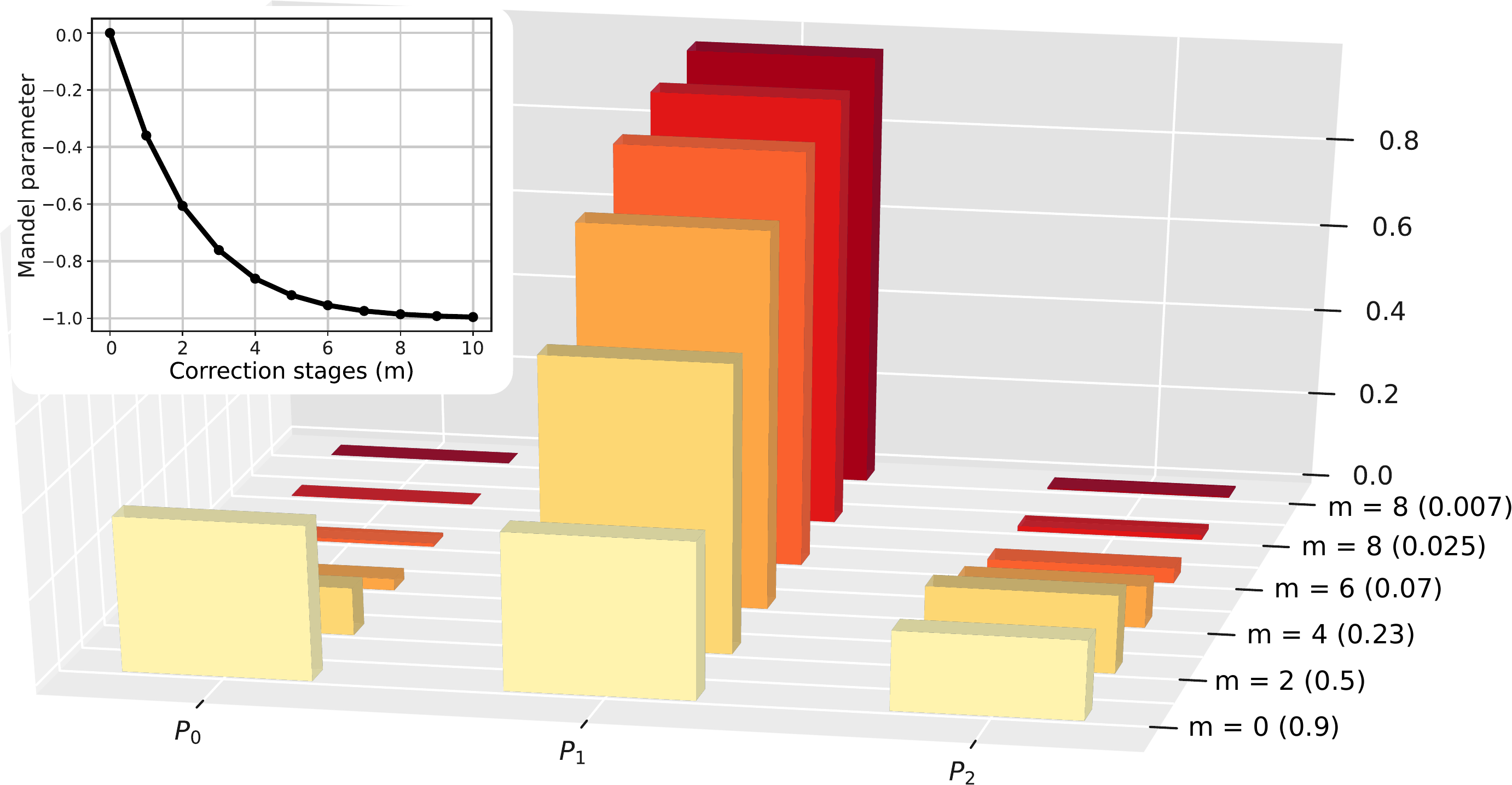}
\caption{Zero, single and multiple emission probabilities for different number of stages $m$, calculated for $\mu^m _{opt}$ values (indicated in brackets). As the chain of correcting stages increases, the probability distributions approach that of a number state. Inset:  $Q_M$ as a function of the amount of correction stages.}
\label{fig:histo}
\end{figure}

The addition of correcting stages increases the total temporal window or correction time ($\Tau$), thus increasing the probability of actually detecting a photon even with low input photon rates. As a consequence, the multiphoton output rate does not increase since the \emph{single} temporal detection window length remains fixed. The effect of multiple correcting stages can be further demonstrated by calculating several terms of the photon output probability distribution, $\mathcal{P}_n$. Figure \ref{fig:histo} shows the discrete photon number distribution for a varying number of correcting stages. These photon probabilities are calculated at the optimum input values $\mu_{opt}$. It can be seen that for increasing values of $m$ the MSPS output approaches a single photon number state. 

A quantitative comparison of this source state and the photon number state can be performed by calculating the Mandel parameter ($Q_M$) in the optimized conditions:
\begin{equation}
\label{ec:mandel}
Q_M= \frac{\Delta ^2 n -\langle  n \rangle }{ \langle n \rangle}=\frac{1}{\langle n \rangle}\sum_{n=0} (n-\langle  n \rangle)^2 \: \text{P}_{n}(\mu_{opt})- 1
.
\end{equation}

For number states ($\Delta ^2 n =0$), $Q_M =-1$. Inset in figure \ref{fig:histo} shows the Mandel parameter for an increasing pile-up of correction stages. As $m$ increases, the variance in the output photon number decreases, ideally achieving values of $Q_M \approx - 0.99$, for $m=10$.

Even though such analysis of a MSPS scheme shows promising results in generating single photon number states that are synchronized with an external clock, it is still an idealized situation, where all the components are considered lossless and that operate with 100\% efficiency. Real life implementations are imperfect and necessarily imply losses. Within the next sections, we present a study of the source performance in the presence of attenuation in optical components and realistic detectors that feature spurious counts and non-ideal efficiency.   

\section{Performance with imperfect devices}
\label{sec:imperfect}

The experimental implementation of any proposed setup implies a degradation in the ideal performance due to the use of real devices. Thus, to properly study the presented source, we need to include these effects in the description. 
The imperfections of this scheme can be classified in three main groups. First, a transmission of the heralding branch can be defined ($e_h$), that takes into account the detector's efficiency and the linear losses in the optical components. These losses will affect the rate of heralding photons being detected, hence reducing the net efficiency of the whole source. State-of-the-art detectors with superconducting technologies present efficiencies as high as 93~\% \cite{transition-edge,reviewMigdall,marsili2013detecting}.

Regarding the same branch, the detector dark counts (R$_{dark}$) must be taken into account: any spurious detection will also act as a heralding photon, triggering the re-routing system with no real correlated photon present in the signal branch. 

Finally, unavoidable losses originate in the optical components that comprise the signal path; specifically the air-to-fiber coupling and the dichroic mirror (that can be summarized in one transmission coefficient $e_{s}$) and, most importantly, the optical switches used for the temporal multiplexing of the heralded photons ($e_{sw}$). The switch insertion loss (IL) is the most critical one, since it strongly limits the amount of correcting stages tolerated by the system. Currently, standard switches based on transparent electro-optic ceramics have an insertion loss of 1 dB or less \cite{switches}.

The previous theoretical model for the source photon number statistics can be adapted to include the presence of the above described losses. In what follows we derive the expression for the $n$-photon output probability with imperfect devices in a constructive way. Taking into account the transmission of the heralding branch only, the photon distribution can be expressed as:

\begin{equation}
\label{ec:p_transmission}
\begin{split}
\mathbf{P}_n(\mu ) = & \big(1-P_0(\mu _T e_h)\big)\frac{P_n(\mu)(1-(1-e_h)^n)}{\sum _j P_j(\mu)(1-(1-e_h)^j)} + \\
& \big( P_0(\mu _T e_h)\big)\frac{P_n(\mu)(1-e_h)^n}{\sum _j P_j(\mu)(1-e_h)^j} ,
\end{split}
\end{equation}
where the first term accounts for $n$ photons at the exit given that a heralding detection occurred, and the second term corresponds to a situation with $n$ photons at the exit but no successful announcement due to total loss of photons on the heralding branch. Each term can be understood as follows: the fist factor determines if there has been an heralding photon emitted and detected ($1-P_0(\mu_Te_h)$) or not ($P_0(\mu_Te_h)$). The second factor corresponds to the probability of having $n$ photons in a single temporal window ($P_n(\mu)$) multiplied by the probability of having at least one heralding click ($1-(1-e_h)^n$) in the successful case or, on the contrary, no clicks ($1-e_h^n$).

Adding dark counts produces the effect of shortening the total correcting temporal window, as a result of a spurious detection. Given the dark count rate, the probability of such event in a \emph{single} temporal window can be written as: $P_{dark}=1-e^{-R_{dark}\Delta t_0}$. Thus, the photon distribution is:
\vspace{-2mm}
\begin{equation}
\label{ec:p_dark}
\begin{split}
\mathbb{P}_n(\mu ) = & \sum _{l=1}^{2^m}(1-P_{dark})^{l-1}P_{dark} \mathbf{P}_n(\mu \,|\, l \: \text{windows})+ \\
& (1-P_{dark})^{2^m}\mathbf{P}_n(\mu \,| 2^m \text{windows}).
\end{split}
\end{equation}

Here, $\mathbf{P}_n(\mu \,|\: l \: \text{windows})$ corresponds to the probability defined in \eqref{ec:p_transmission} but with a total synchronization interval $\Tau_l = l \times \Delta t_0$. The first term accounts for the case when a dark count is detected in the $l$-th window (the probability of such event is $(1-P_{dark})^{l-1}P_{dark}$), multiplied by the probability of equation \eqref{ec:p_transmission} with the correspondingly shortened synchronization interval. This is summed over all possible single detection windows in $\Tau$. The second term corresponds to the case when no dark counts are registered in $\Tau$ ($(1-P_{dark})^{2^m}$): in that case the output probability is exactly that of equation \eqref{ec:p_transmission}. 

Finally, the loss in the signal (heralded) branch is taken into account: the overall transmission can be expressed as $e_{s}^{tot} = e_{s}\times (e_{sw})^{m + 1}$,
which adds a binomial factor to the probability stated in \eqref{ec:p_dark}:

\vspace{-2mm}
\begin{equation}
\label{ec:p_tot}
\mathcal{P}_k(\mu ) =  \mathbb{P}_n(\mu)\times B(k,N = n,p = e_s^{tot}),
\end{equation}
where $B(k,N = n,p = e_s^{tot})$ is the binomial coefficient for $k$ successes, having tried $N = n$ times, with a success probability of $p = e_s^{tot}$. The probability of having $k$ photons at the exit of the source for a input mean photon number $\mu$, taking into account all losses of the system is given by $\mathcal{P}_k(\mu )$. 

\begin{figure}[h]
\centering
\includegraphics[width=0.48\textwidth]{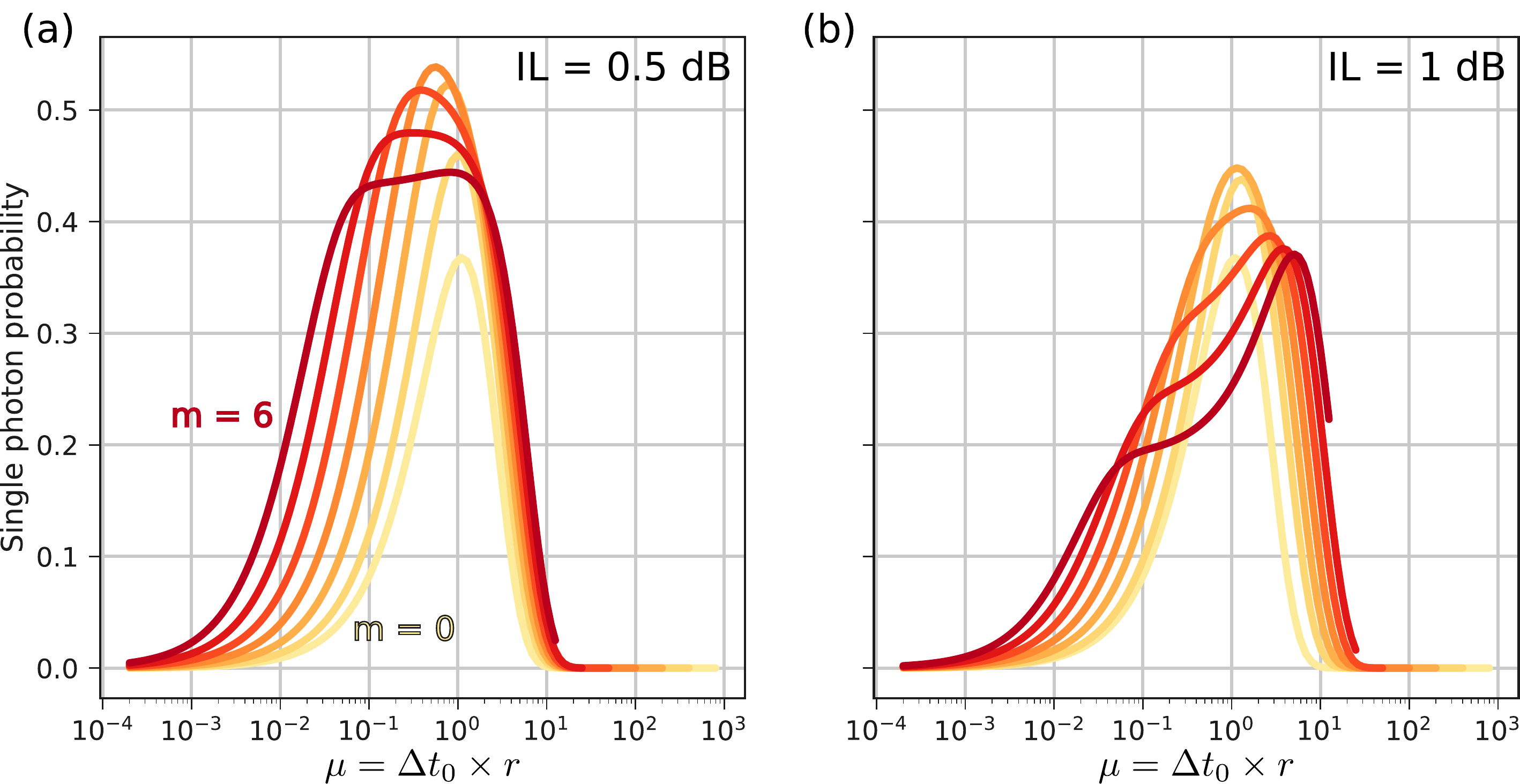}
\caption{Probability of a single photon as a function of the input mean photon number for $e_{h} = 0.85$ and $e_{s} = 0.9$ and two different switch IL values: (a) shows the result for $0.5$~dB, and (b) for $1$~dB. Each curve corresponds to $m$ correction stages. As $m$ increases, so does the loss present in the system. Both figures share the y axis.}
\label{fig:p1_imperfect}
\end{figure}

In order to visualize and compare these results with the ideal situation, we show the single photon probability as a function of $\mu $ (Fig. \ref{fig:p1_imperfect}), for two different switch IL values: $e_{sw} = 0.5$~dB and $e_{sw} = 1$~dB. For these calculations, the transmission of the idler and signal paths were set to $e_{h} = 0.85$ and $e_{s} = 0.9$.

For $e_{sw} = 0.5$~dB, there are several stages that clearly show an improvement from the poissonian result of a weak coherent pulse source ($m = 0$). An optimal input mean photon number can still be identified until $m = 5$, but the maximum single photon probability starts to decrease for $m > 3$. For $1$~dB switch IL, in the condition of maximum probability, only $m = 1$ and $m = 2$ correction stages outperform the results for $m = 0$. For higher numbers of correcting stages, a maximum can still be found, but this corresponds to a condition in which attenuation and multi-photon emission dominate the system output.

\begin{figure}[h]
\centering
\includegraphics[width=0.48\textwidth]{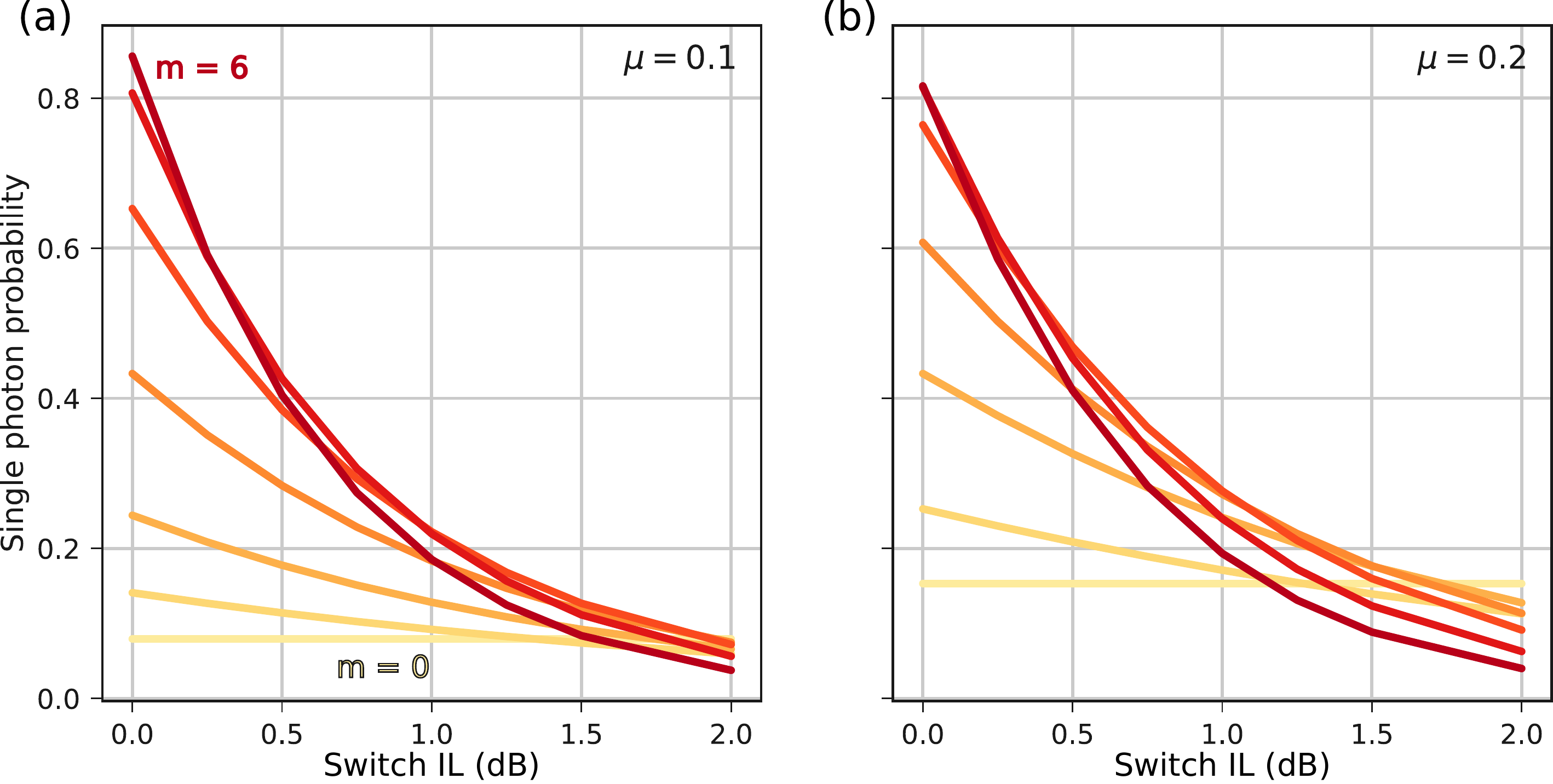}
\caption{Single photon probability for increasing switch IL (in dB), for different lengths of correction stages (each plot line) and a fixed input mean photon number. (a) corresponds to $\mu = 0.1$ and (b) to $\mu = 0.2$.}
\label{fig:sw_loss}
\end{figure}

Such condition is not desirable for quantum information processing, since 
it merely corresponds to the maximization of the single photon probability, whereas the multiple photon emission probability $\mathcal{P}_{\geq2}$ can reach unacceptably high rates. It is therefore useful to explore the behavior of the source for lower input photon rates. For $\mu < \mu _{opt}$, the addition of the multiplexing system increases the probability of a single photon, improving the system overall performance. Thus, the value of $\mu$ that maximizes the effect of adding correction stages depends on the loss present in the experimental arrangement. Figure \ref{fig:sw_loss} shows the single photon probability achieved for a given input mean photon number [$\mu = 0.1$ (a) and $\mu = 0.2$ (b)], as a function of the switch IL for several lengths of correction stages. 

Furthermore, to quantify the multiphoton contribution, the Signal-To-Noise ratio of the source can be calculated:
\begin{equation}
\text{SNR}(\mu ) = \frac{\mathcal{P}_1(\mu )}{\mathcal{P}_{\geq2}(\mu )}.
\end{equation}

\begin{figure}[h]
\centering
\includegraphics[width=0.48\textwidth]{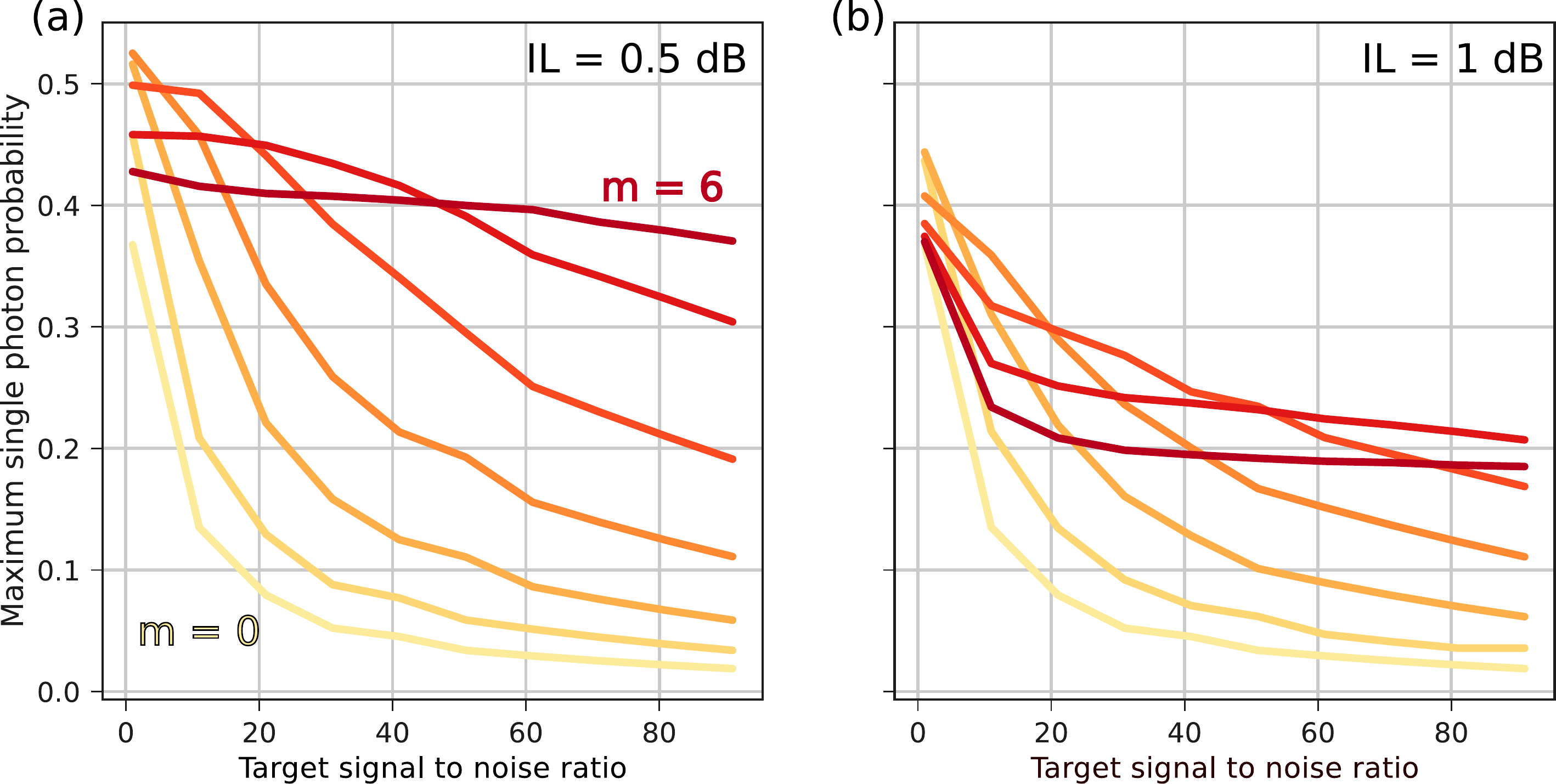}
\caption{Maximum single photon probability as a function of the Signal-To-Noise ratio target, for several amounts of correction stages. Results of two different switch IL conditions are shown: (a) for 0.5 dB per switch, and (b) for 1 dB per switch.}
\label{fig:snr}
\end{figure}

In general, to guarantee a small rate of multiphoton pulses, a SNR threshold can be established. Figure \ref{fig:snr} shows the maximum single photon probability obtained for different amounts of correction stages, provided a minimum of Signal-To-Noise ratio. Depending on the tolerance on multiphoton events that can be accepted, the amount of correction stages that is convenient can be calculated, thus defining the maximum single photon probability that can be achieved in that condition. Figure \ref{fig:snr} shows that for high values of target SNR, the addition of correction stages becomes more significant. 

As the switch IL increases, the beneficial effect of adding correction stages becomes neutralized by the attenuation introduced by the network. 
The ideal working condition must be decided upon the specific application of the source and the loss of the available switches. Once a target SNR is defined, the convenient amount of correcting stages and input photon rate can be calculated. 

For example, in the case of $0.5$ dB loss and an application where a SNR close to 50 is required, a remarkable improvement from the weak coherent pulse source can be achieved using $m =4$ stages. For example, for an input mean photon number of $\mu=10^{-1}$ per detection window, the addition of up to 4 correction stages rises the single photon probability from $\approx0.08$ of a weak coherent pulse source to $\approx0.4$, while the Signal-To-Noise ratio doubles from $\approx22$ to $\approx44$.
For the particular choice of temporal parameters of the source described herein, the synchronizing clock runs at 31 MHz. 
In turn, for high loss devices [figures \ref{fig:p1_imperfect}, \ref{fig:sw_loss} and \ref{fig:snr} (b)] the probability of heralding a zero-photon output is strongly dependent on the addition of correcting stages, and the single-photon output is limited, but the Signal-to-Noise ratio can still be improved: a value of $\mathcal{P}_1=0.2$ can be obtained directly from the SPDC source ($m=0$), or with the aid of 4 correcting stages ($m=4$). The device lowers the multiphoton output probability, by requiring lower input photon rates, thus increasing the SNR from 10 to 50. 

These results represent an improvement with respect to the weak coherent pulse case. However, it is worth to note that further enhancement on the ``single photon output'' could be obtained by using a combination of time multiplexing schemes: the device discussed in this work can be used as the photon source for time multiplexing arrangements based on a storage cavity, like the scheme studied in \cite{kwiat_new, Kwiat_old}, which requires a periodic pulsed input. Such combination would have the advantages of both increasing the single photon probability (compared to that of each multiplexing technique alone) and allowing the whole device to be used with a continuos-wave laser as pump.

\section{Conclusions}
\label{sec:conclusions}

Based on the proposed setup for a single photon source with a time multiplexing techniques, we have developed a theoretical model for the photon statistics that takes into account real life implementation effects such as component loss, detector efficiency and dark counts.

The single photon source is based on a time multiplexing system consisting of an actively switched fiber delay network with a binary division of the path lengths. The selected delay is set using the arrival time of the idler photon with respect to a clock tick. This source can synchronize the emission of a single photon originated in a continuously pumped SPDC source to an external time signal. The idealized single photon output probability depends, in principle, only on the size of the fiber network and the input power, although in a real-life implementation the IL of the optical switches reduces the performance of the device. Our model accounts for such losses and also for spurious events like dark counts, and optimum parameters for sub-poissonian output statistics can be found under realistic working conditions.  Interestingly, a 1 dB IL for the switching devices seems to be a limiting value to observe an enhancement on the source behavior.

Finally, this multiplexing technique has the advantage that it can be used as an improved input for other time multiplexing techniques that require periodic pumps. This combination of time multiplexing techniques would represent an improvement over the performance of each technique alone.

\begin{acknowledgements}
We acknowledge financial support from CONICET and ANPCyT funding agencies, and the \mbox{PIDDEF} program from MINDEF. 
\end{acknowledgements}

%


\end{document}